\documentclass[%
 reprint
 amsmath,amssymb,
 aps,
 prl,
]{revtex4-1}

\usepackage{graphicx}
\usepackage{dcolumn}
\usepackage{bm}
\usepackage{amssymb}
\usepackage{epstopdf}
\usepackage{color}

\begin{document}

\title{Computing the Threshold of the Influence of Intercellular Nanotubes on
 Cell-to-Cell Communication Integrity}

\author{Dragutin T. Mihailovi\'c}
\affiliation{Faculty of Agriculture, University of Novi Sad, Serbia}
\email{guto@polj.uns.ac.rs}

\author{Vladimir R. Kosti\'c}%

\author{Igor Bala\v{z}}%

\author{Darko Kapor}%

\affiliation{Faculty of Sciences, University of Novi Sad, Serbia}%

\date{\today}

\begin{abstract}
We examine the threshold of the influence of the tunneling nanotubes (TNTs) on the cell-to-cell communication integrity. A deterministic model is introduced with the Michaelis-Menten dynamics and the intercellular exchange of substance. The influence of TNTs are considered as a functional perturbation of the main communication and treated as the matrix nearness problems. We analyze communication integrity in terms of the \emph{pseudospectra} of the exchange, to find the \emph{distance to instability}. The threshold of TNTs influence is computed for Newman-Gastner and Erd\H{o}s-R\'enyi gap junction (GJ) networks.
\end{abstract}

\maketitle

\begin{figure}[b]
\includegraphics{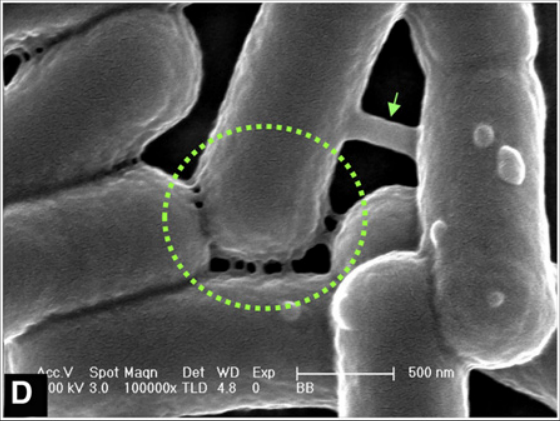}
\caption{\label{fig:prepTNT} Intercellular TNTs between neighboring cells. A field of cells with a cluster of smaller TNTs (highlighted by a dashed circle) and a more pronounced larger tube (indicated by an arrow). Reprinted with permission from \cite{Dubey2011}.}
\end{figure}

Maintaining the functional integrity of cell-to-cell communication in multicellular systems is one of the prerequisites for achieving functionality in them \cite{Levin2006, Pikovsky2001, Arenas2008, Chen2003, Mihailovic2014}. In eukaryotic cells, intercellular communication is primarily mediated locally through gap junctions (GJs) and synapses; however, recent reports demonstrate the existence of a network of intercellular membrane nanotubes enabling long-distance communication \cite{Ramirez1999, Onfelt2004, Rustom2004}. Also in prokaryotes they enable interspecies communication and share of antibiotic resistance \cite{Dubey2011, Ficht2011}. It has been shown that in both cases these tunneling nanotubes (TNTs) can facilitate cell-to-cell communication and intercellular transfer of cytoplasmic molecules, organelles and viruses  \cite{Rustom2004, Belting2008, Sowinski2008, Bukoreshtliev2009}. Existence of clusters of TNTs enables formation of complex cellular networks with both local and long-distance interactions based on membrane continuity between TNT-connected cells (Fig.~\ref{fig:prepTNT}). Empirical evidence indicate that they can have important role in many pathophysiological processes, like in activation of natural killer cells, regulation of osteoclastogenesis or in tumor formation and growth \cite{Chauveau2010, Takahashi2013, Lou2012}. In prokaryotes they can play important part in transferring virulence from pathogenic to non-pathogenic bacteria \cite{Dubey2011} These findings indicate the need to systematically explore how perturbations in communication, induced by existence of clusters of TNTs, can influence integrity of intercellular communication.

In this Letter we explore how the substance exchange through TNTs affects the functional stability of a multicellular system. We focus on two issues: (i) whether TNTs can either stabilize or destabilize intercellular communication governed by GJs? and (ii) how to determine the threshold of TNTs influence that shifts stable GJs communication to the unstable one? Therefore, we introduce a model of the substance exchange in a multicellular system, represented by ordinary differential equations, where cell-to-cell communication is mediated by both the GJs and TNTs while metabolic processes in cell follow Michaelis-Menten dynamics. In this model the GJs function governs the time evolution of the intercellular network while the TNTs function simulates the exchange mediated by the TNTs including a scaling parameter for that mediation. We consider the influence of TNTs as a functional perturbation of the main communication mediated by GJs. To determine the threshold for which the multicellular system remains stable, despite TNTs influence, we compute the pseudospectra of the exchange to find the \emph{distance to instability} \cite{Higham1989}.

\paragraph{General model dynamics and pseudospectra.}
To investigate how TNTs affect the stability of the intercellular communication network, we model network dynamics as:

\begin{equation}\label{model1}
\dot x(t) = \Psi(x(t)):=\Phi(x(t))+\xi \Theta(x(t)),
\end{equation}
where $x=[x_1,x_2,\ldots,x_n]^T$. Here $x_i(t)\in[0,1]$ is the concentration of molecules and ions in the cell $i \in N :=[1, 2, \ldots, n]$, $\Phi:R^{n}\to R^{n}$ is a GJs function that governs time evolution of the intercellular network, while  $\Theta: R^{n}\to R^{n}$ models the exchange mediated by TNTs and $\xi>0$ is a scaling parameter for that mediation. TNTs are dynamic structures whose formation and decomposition is very sensitive to both intra- and extracellular factors, with lifetimes ranging from several minutes up to 4-5 hours \cite{Lou2012, Bukoreshtliev2009, Rustom2004}. Their short lifetime suggests that they can promote unstable, transient cell-to-cell communication, in contrast to more stable GJs \cite{Goodenough2009}. Since many questions remain unanswered about how cargo is transported through TNTs \cite{Davis2008} we consider their effect on the model dynamics as an uncertainty described by $\Theta$. The system (\ref{model1}) is generally a nonlinear one whose stability is typically investigated at the equilibrium state $x^\star\in R^{n}$ as the local asymptotic stability, where $x^\star$ is a state vector such that $\Psi(x^\star)=0$. An equilibrium state $x^\star$ is locally asymptotically stable if there exists $\epsilon>0$ such that for every $x(0)$ that is in $\epsilon$ neighbourhood of the equilibrium $x^\star$ (i.e., $\|x^\star-x(0)\|< \mu $), it holds that $\displaystyle{\lim_{t\to\infty}}\|x(t)-x^\star\|=0$. This stability property is characterised by the spectra of the Jacobian  matrix $A=\left[\frac{\partial\Psi_i}{\partial x_j}(x^\star)\right]$ of (\ref{model1}) in the state $x=x^\star$, which  can be written as $A=\widehat A + \xi \Delta$, where $\widehat A=[\frac{\partial \Phi_i}{\partial x_j} (x^\star)]$ corresponds to the exchange through GJs that will be named as the observable part. The term $\widehat \Delta=[\frac{\partial \Theta_i}{\partial x_j} (x^\star)]$ is determined by the transport through TNTs that is generally unknown. Thus, we call it the unobservable part. Furthermore, for $\xi>0$, to reflect the scale of TNT mediation we assume that uncertainty parameters are unit scaled in the chosen matrix norm, i.e. $\|\widehat \Delta\|=1$.

To investigate how TNTs can influence stability of communication we will analyze sensitivity of the spectrum of the observable matrix $\widehat A$ upon perturbation $\xi \Delta$, using the concept of matrix pseudospectra \cite{Trefethen2005}. Given a matrix $\widehat A\in \mathbb{C}^{n,n}$ and $\varepsilon>0$, the $\varepsilon$ -pseudospectrum of a matrix $\widehat A$, denoted by  $\Lambda_{\varepsilon}( \widehat A)$, is composed of all eigenvalues of matrices which are "\emph{$\varepsilon -$close}" to $\widehat A$:
$\lambda\in\Lambda_{\varepsilon}( \widehat A)$ if and only if there exists $x  \in \mathbb{C}^n \setminus {0}$ and  $\widehat\Delta  \in \mathbb{C}^{n,n}$ such that $\|\widehat\Delta\| \leq \varepsilon$ and $(\widehat A+\widehat\Delta)x=\lambda x$, i.e. $\Lambda_{\varepsilon}(\widehat A) = \bigcup_{\|\widehat\Delta\|\leq\varepsilon}\Lambda(\widehat A+\widehat\Delta)$. Consequently, we use $\varepsilon$-pseudospectrum to establish spectral properties that are robust under matrix perturbations bounded in a given norm $\|\cdot\|$ by the parameter $\varepsilon>0$. The measure of robustness of the stability of the observable matrix $\widehat A$ is defined as the largest $\varepsilon>0$ such that $\Lambda_\varepsilon(\widehat A)\subset \mathbb{C}^{-}$, and is known as the \emph{distance to instability} \cite{Higham1989}. Since the computation of such value $\varepsilon>0$ requires solving a nonconvex optimization problem, numerical algorithms have to be used \cite{Byers1988, Trefethen2005, He1999, Freitag2011, Gugliemi2015}. Computing the spectra $\Lambda(\widehat A)$  of the observable part of the Jacobian, corresponding to GJs interactions and intracellular metabolic processes, we can determine the expected stability $\Lambda(\widehat A)\subseteq \mathbb{C}^{-}$ or instability $\Lambda(\widehat A)\not\subseteq \mathbb{C}^{-}$ of the substance fluxes.  However, the unobservable part of cellular communication can change this spectral property and lead to the different dynamics of the network since $\Lambda(A)\subset\Lambda_{\xi}(\widehat A)$. More precisely, assuming that $\Lambda(\widehat A)\subseteq \mathbb{C}^{-}$, then $
\Lambda(A)\subset\Lambda_{\xi}(\widehat A)\subseteq \mathbb{C}^{-}$  if and only if $\xi<\nu^{-}_{\widehat A}$, where $\nu^{-}_{\widehat A}$ denotes the distance to instability of $\widehat A$. Therefore, if the TNTs influence is dominated by the distance to instability of the observable matrix $\widehat A$, we can safely conclude that the mediation of TNTs does not change the stability of (\ref{model1}), while for the values $\xi\geq \nu_{\widehat A}^{-}$ we cannot do that. Otherwise, assuming that $\Lambda(\widehat A)\not\subseteq \mathbb{C}^{-}$, then $\Lambda(A)\not\subseteq \mathbb{C}^{-}$ if and only if $\xi<\nu^{+}_{\widehat A}$,
where $\nu^{+}_{\widehat A}$ denotes the distance to stability of $\widehat A$. Thus, in this case, distance to stability of the observable matrix $\widehat A$ is the critical value of the TNTs influence for which the system can change the unstable behaviour to the stable one. Therefore, the integrity of GJ communication (either stable or unstable one) under the influence of TNTs does not depend only on GJ network's \emph{resilience}, i.e. real part of the least negative eigenvalue of the observable Jacobian matrix. In fact, it remarkably depends on the GJ network structure which can produce non-normal dynamics \cite{Trefethen2005}.

\paragraph{Deterministic model.}
We introduce a deterministic model for substance exchange in a multicellular system which is mediated by GJs and TNTs. Both, GJs and TNTs allow various molecules and ions to pass freely between cells through the channels by the diffusion like process. However, diffusion in cells and between them, known as “anomalous diffusion“, can differ from “classical” one due to spatial inhomogeneity \cite{Nitche1999, Cherstvy2013, Mullineaux2008}. In situation like this, it is suitable to consider the kinetics of substance exchange between cells in terms of an \emph{exchange coefficient} $g_{ij}$ with dimensional unit of inverse time \cite{Nitche2004}. In the simplest case, communication from the cell $i$ to cell $j$ is proportional to the concentrations between the cells. Therefore, we can define the substance exchange between cell $j$ to cell $i$ as $g_{ij} (x_j (t)-x_i (t))+\xi\delta_{ij} x_j (t)$, where $\xi>0$ is a small value that determines the strength of influence of TNTs on communication modeled by uncertainty parameter $\delta_{ij}$. Since exchanged substances play a role in the metabolic processes inside the cell and are released into environment, we introduce parameters $\alpha_i>0$ that describe the rate by which the substance is metabolized by the cell $i\in N$ in time $t$. Since most of metabolic processes follow Michaelis-Menten dynamics, we introduce $\beta_i>0$ as the half-time saturation coefficient for the cell $i$. Therefore, we express the intercellular communication as:

\begin{equation}\label{model2}
\dot x_i(t) = -\frac{\alpha_i x_i(t)}{\beta_i+x_i(t)} + \sum_{j\neq i} g_{ij}\left(x_j(t)-x_i(t)\right) + \xi \sum_{j\in N} \delta_{ij} x_j(t), \quad (i\in N),
\end{equation}

Finally, we restrict to the case of zero equilibrium $x^\star=0$,  when the initial state $x(0)\neq0$ reflects starting distribution of the substance in the network. Consequently, $\widehat \Delta=[\delta_{ij}]$ with $\|\widehat\Delta\|=1$, while $\widehat A=[\frac{\partial \Phi_i}{\partial x_j} (0)]=[\widehat a_{ij}]$ where for $i,j\in N$ $ \widehat a_{ij}= -(\frac{\alpha_i}{\beta_i}+\sum_{j\neq i}g_{ij})$ if $i = j$ or $ \widehat a_{ij} = g_{ij}$ if $i\neq j$. We use such a simple deterministic model because: (i) it is general enough to embrace the main aspects of cell-to-cell substance exchange via GJs and TNTs and (ii) it, without the loss of generality, clearly illustrates the use of pseudospectra in estimating the uncertainty in the model dynamics determined by the substance exchange through TNTs.

Depending of the system's quantity or property which is examined, we chose the norm in which distances are measured. Here we discus two cases. First, that is the maximal substance concentration in the network with norm set to infinity norm. In that setting, for communication mediated exclusively by GJs, i.e. $\xi=0$, using pseudospectral localization \cite{Kostic2015a},  we can conclude that zero equilibrium point of (\ref{model2}) is exponentially stable one, and that for all $\xi\geq0$, the flow of (\ref{model2}) $x(t)$ satisfies $\|x(t)\|_\infty\leq e^{-M+\xi}\|x(0)\|_\infty$ for $ t\geq 0$, where $M=\min_{i\in N}\frac{\alpha_i}{\beta_i}>0$. In other words, for $\xi<M$, the maximal substance concentration in individual cells exhibits an exponential decay without the initial growth. This implies  that the system maintains its communication integrity. Therefore, the  obtained constant $M>0$ is a lower bound of the distance to instability of $\widehat A$ in $\|\cdot\|_\infty$. Second, to check the behavior of the network in terms of total square deviation from the equilibrium state, the norm has to be set to the Euclidian one. Therefore, to compute the threshold of TNTs influence, in that setting, we apply the algorithm developed by Kosti\'c et al. \cite{Kostic2015b}, while the pseudospectral portraits are generated using the EigTool \cite{Wright2002}.

\begin{figure*}
\includegraphics [width=0.8\textwidth, keepaspectratio=true] {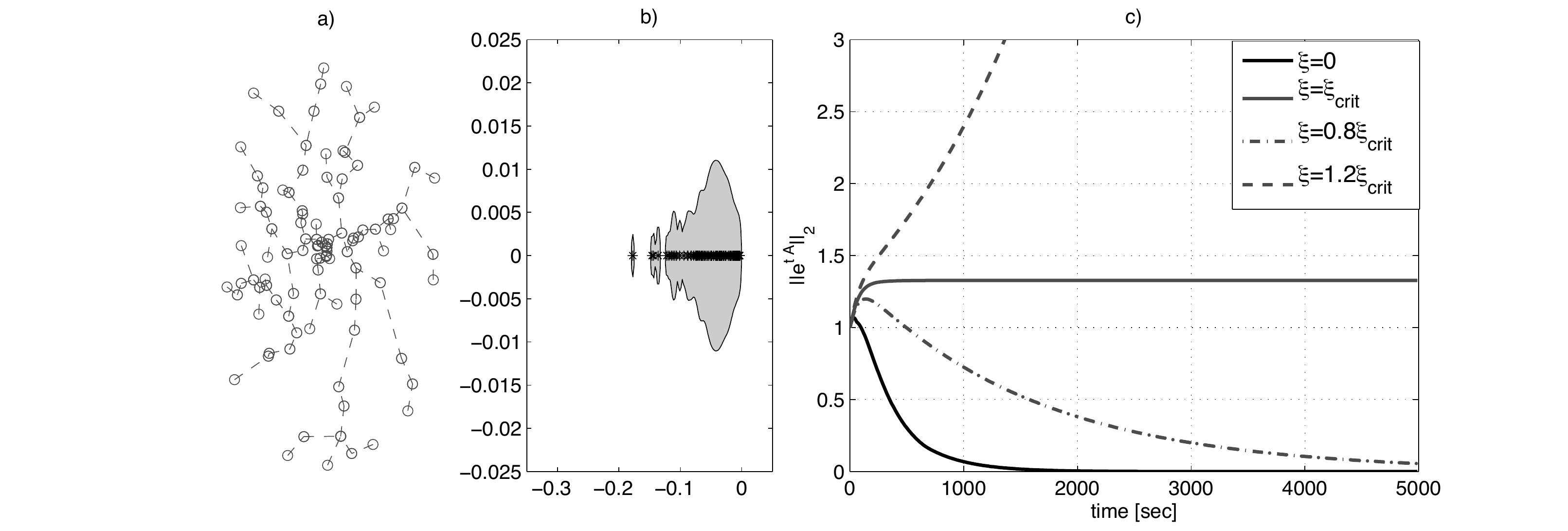}
\caption{\label{fig:NewmanGastner} Distance to instability computed for $100$-cell GJ network constructed as the Newman-Gastner spatial network. (a) Graphical image of that network; (b) pseudospectral portrait of its Jacobian matrix $\widehat A$ (asterisks mark eigenvalues of $\widehat A$ $(\xi=0)$, while the shadowed area represents all the possible locations of eigenvalues for the full network Jacobian $A$, when GJs and TNTs are included ($\xi = \xi_{crit}$ where $\xi_{crit}=\nu_{\widehat A}^{-}$ is a threshold value)); (c) transient growth of substance concentration within cells from the initial state measured in the Euclidian norm $\|e^{t A}\|_2$, due to non-normality of the GJs Jacobian matrix \cite{McCoy2013} for the following cases: $\xi = 0$ (solid black), $\xi = \xi_{crit}$ (solid gray), $\xi = 0.8\xi_{crit}$ (solid dashed-dotted) and $\xi = 1.2\xi_{crit}$ (solid dashed). Note that peak of concentration and duration of concentration decay depends on the scenario used as seen in Figs 2c-4c. The computed values are $\xi_{crit} = 2.38\cdot10^{-3}$, and $M=4.47\cdot10^{-5}$.}
\end{figure*}

\begin{figure*}
\includegraphics [width=0.8\textwidth, keepaspectratio=true] {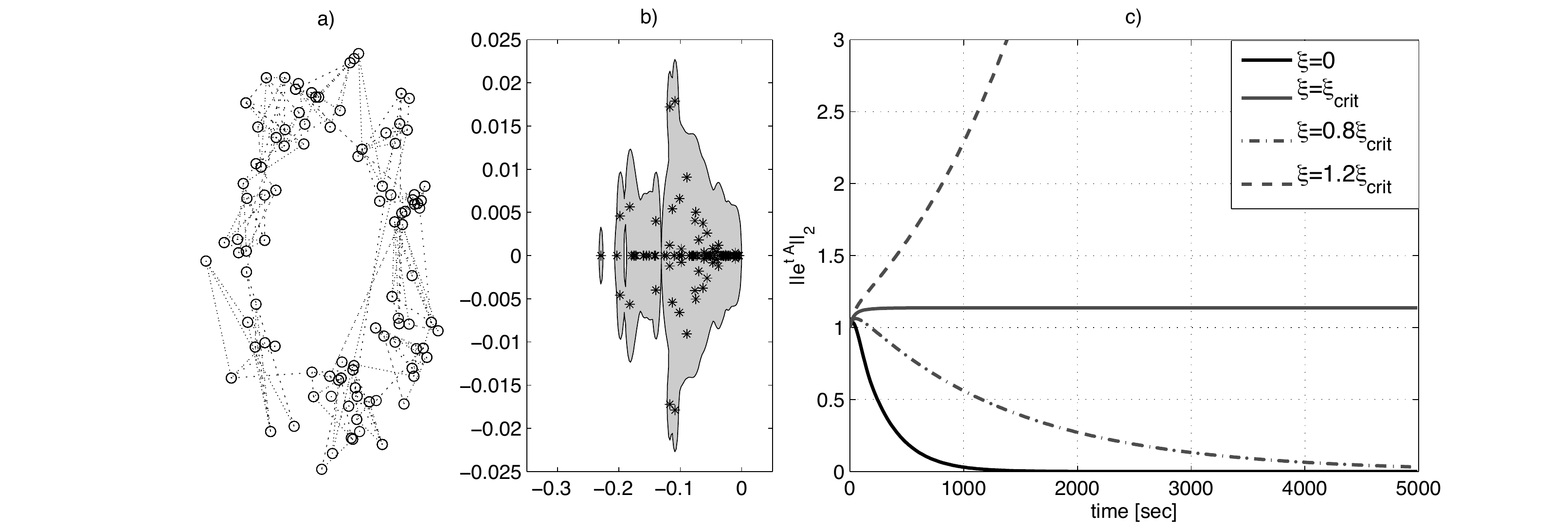}
\caption{\label{fig:Erdos-Renyi} The same as in Fig. \ref{fig:NewmanGastner} but for Erd\H{o}s-R\'enyi modular network. The  computed values are $\xi_{crit} = 3.15\cdot10^{-3}$, and $M=6.30\cdot10^{-5}$.}
\end{figure*}

\begin{figure*}
\includegraphics [width=0.8\textwidth, keepaspectratio=true] {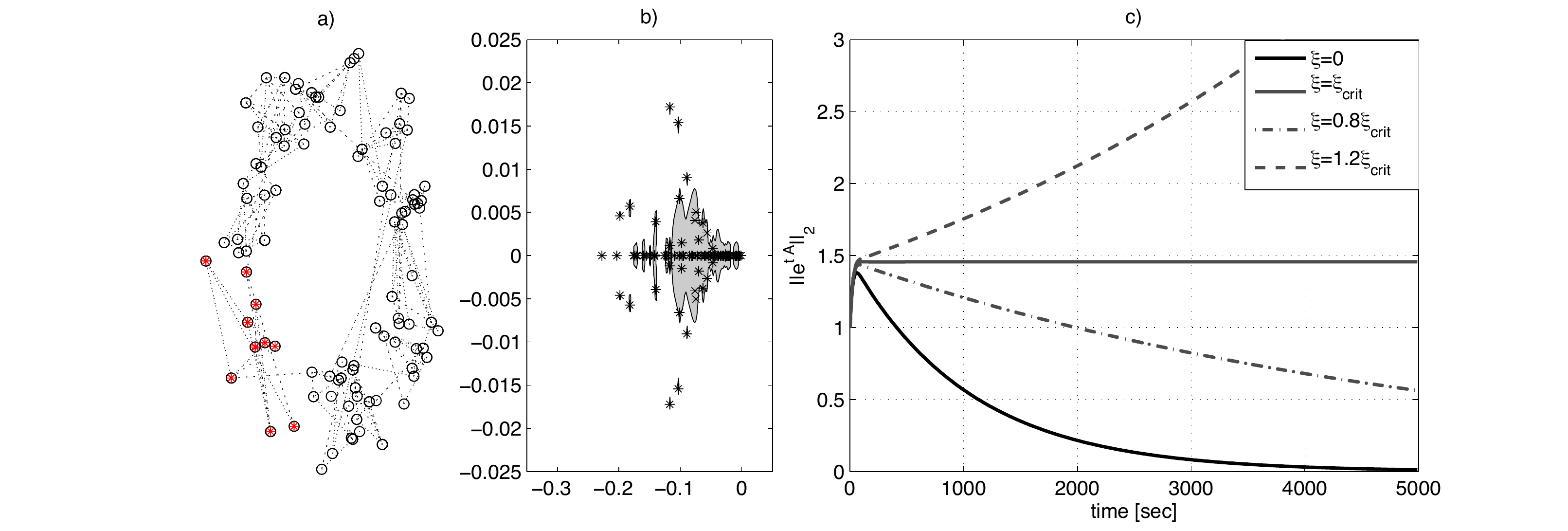}
\caption{\label{fig:Erdos-Renyi-path} The same as in Fig. \ref{fig:NewmanGastner} but for Erd\H{o}s-R\'enyi random network simulating a pathological state of the system (colored circles show nodes with altered capacity to receive signals). The  computed values are $\xi_{crit} = 6.55\cdot10^{-4}$, and $M=6.30\cdot10^{-5}$.}
\end{figure*}

\paragraph{Results and discussion.}
We consider a few realistic scenarios based on the simple deterministic model of $100$-cell network. In all test cases we compute lower bound of threshold of TNTs influence in infinity norm ($M$), the exact value in Euclidean norm ($\nu_{\widehat A}^{-}$) and construct pseudospectral portrait with transient plot. For GJ communication we use: (i) spatially distributed Newman-Gastner network with weight parameter set to $0.001$ \cite{Gastner2006} and (ii) Erd\H{o}s-R\'enyi modular network \cite{Erdos1959} with ten clusters connected wit $0.03$ overall probability of the attachment and proportion of links within modules set to $90\%$, graphically depicted in Figs. \ref{fig:NewmanGastner}a, \ref{fig:Erdos-Renyi}a  and \ref{fig:Erdos-Renyi-path}a. For values of physiological parameters in simulations, we assume that all cells in the population are of the same type and therefore have the same time scales for metabolizing the substance. Therefore, for each $i = 1,\dots, 100$, we take  $\beta_i$  randomly on an uniformly distributed interval $[0.9, 1.1]$. On the other hand, the saturation constants {$\alpha_i$} (having the same order of value as the exchange coefficients), that are of the same order as the exchange coefficients, can differ more significantly, and are chosen randomly from $[0, 0.01]$ with the uniform distribution. Finally, following \cite{Mullineaux2008}, for exchange coefficients $g_{ij}$ we use random values from the interval $[0, 0.05]$ with the uniform distribution.

We compute the critical pseudospectra of the GJs Jacobian matrix $ \widehat A$  of the network (Figs. \ref{fig:NewmanGastner}b, \ref{fig:Erdos-Renyi}b and \ref{fig:Erdos-Renyi-path}b).  Here, the term critical pseudospectra stands for the fact that $\varepsilon=\nu_{\widehat A}^{-}$  which is the computed threshold of TNTs influence. The shadowed area indicates how the system is  sensitive to changes in cell communication determined by TNTs. To demonstrate how formation of TNTs can affect network dynamics, we compute first order approximated behavior of (\ref{model2}), measured in Euclidean norm, i.e. $\|e^{t A}\|_2$  (Figs. \ref{fig:NewmanGastner}c, \ref{fig:Erdos-Renyi}c and \ref{fig:Erdos-Renyi-path}c), for the following cases. \emph{Case $1$}, when cell-to-cell communication takes place only through GJs $(\xi = 0)$. Then the system, after passing through short transient interval, is reaching the stability either faster (Figs. \ref{fig:NewmanGastner}c and \ref{fig:Erdos-Renyi}c) or slower (Figs. \ref{fig:Erdos-Renyi-path}c). The corresponding curves are depicted by the solid black lines. \emph{Case $2$} (critical case), when $\xi = \xi_{crit}$ and $\widehat \Delta$ is the worst arrangement of TNTs that move eigenvalues of $A$ to the imaginary line (\emph{marginal instability}). Such $\widehat \Delta$ is constructed via suitable singular vectors \cite{Kostic2015b}. From Figs. \ref{fig:NewmanGastner}c, \ref{fig:Erdos-Renyi}c and \ref{fig:Erdos-Renyi-path}c (solid gray line) it is seen that the substance exchange in the system is in the state of an oscillating mode, waiting to start towards either stability or instability. \emph{Case 3} (case of stability) when $\xi < \xi_{crit}$ (here $\xi = 0.8\xi_{crit}$). In this case the system maintains the communication integrity (gray dashed-dotted line in Figs. \ref{fig:NewmanGastner}c, \ref{fig:Erdos-Renyi}c and \ref{fig:Erdos-Renyi-path}c). \emph{Case 4} (case of instability), when $\xi > \xi_{crit}$ (here $\xi = 1.2\xi_{crit}$). Correspondingly, the system is communicationally disintegrated (gray dashed line in the Figs. \ref{fig:NewmanGastner}c, \ref{fig:Erdos-Renyi}c and \ref{fig:Erdos-Renyi-path}c).

The situation of disrupted cell-to-cell communication can cause numerous diseases. For example, inborn cardiac diseases are among the most frequent congenital anomalies and are caused by mutations in genes that form GJs \cite{Salameh2013}. Therefore, it is crucial to investigate stability of intercellular communication and determine possible thresholds for disruption of cell-to-cell communication integrity. To investigate a possible influence of TNTs in the case of disrupted communication we simulate pathological situation by modifying the Erd\H{o}s-R\'enyi modular network as follows. In this network arrangement only one module (colored circle in Fig. \ref{fig:Erdos-Renyi-path}a) exhibits a cascade degradation of their capacity to receive the substance under exchange, while their capacity to send it in the fixed network flux direction (corresponding to the node enumeration) remains the same. More precisely, in the original realization of Erd\H{o}s-R\'enyi modular network we take $0.1g_{ij}$, instead of $g_{ij}$ for $i=1\ldots100$ and $j=1\ldots10$. In the example we create, when only one module is disrupted, pseudospectral portrait (Fig.\ref{fig:Erdos-Renyi-path}b) shows that sensitivity of the whole network to communication changes is significantly increased, compared to non-pathological state (Fig.\ref{fig:Erdos-Renyi}b). Also, critical level of oscillations in Euclidian norm deviate more from the equilibrium state indicating that formation of TNTs can disrupt the system more easily.

To summarise, the main novelty of this Letter lies in the fact that the uncertainty of TNTs influence to the overall cellular communication can be treated as the matrix nearness problems, i.e. either as distance to instability (treated in this Letter) or distance to stability (reverse problem that will be treated in the follow-up study). We have presented how this concept can provide meaningful insights using the simple deterministic model of cellular networks with asymptotically stable GJs cell-to-cell communication, where TNTs can destabilize the system. The problem is analysed in terms of maximal individual deviation ($\|\cdot\|_\infty$) and total square deviation (Euclidian norm) of the cells substance concentration. In the simulations, the threshold of such TNTs influence is computed using recently developed pseudospectra methods for two standard structures of cellular networks (spatial Newman-Gastner and modular Erd\H{o}s-R\'enyi) modelling healthy and pathological state of the system. The work presented here is a first step towards understanding of the influence of TNTs as uncertainty of the system using matrix analysis and computational methods. Therefore, many open questions remain, targeting, among other ones, the tie between matrix perturbations as stochastic processes, where the pseudospectra combined with the Bregman divergences \cite{Dhillon2007}  can help to reliably estimate the mathematical expectation of the threshold of destabilising/stabilizing intercellular communication.

This work is supported by the Ministry of Education and Science of the Republic of Serbia (Grants III 43007 and 174019) and Provincial Secretariat for Science and Technology Development of Vojvodina (Grants 1136, 1850 and 2010).

\bibliography{apssamp}

\end{document}